\documentclass{optica-article}

\journal{opticajournal}

\setprjcopyright

\articletype{Research Article}

\begin{document}

\title{Equalization of a 10 Gbps IMDD signal by a small silicon photonics time delayed neural network}

\author{Emiliano Staffoli,\authormark{1,*} Mattia Mancinelli,\authormark{1}, Paolo Bettotti, \authormark{1} and Lorenzo Pavesi\authormark{1}}

\address{\authormark{1}Nanoscience Laboratory, Department of Physics, University of Trento, Via Sommarive 14, 38123, Trento, Italy}

\email{\authormark{*}emiliano.staffoli@unitn.it} 



\begin{abstract}
A small 4-channels time-delayed complex perceptron is used as a silicon photonics neural network (NN) device to compensate for chromatic dispersion in optical fiber links. The NN device is experimentally tested with non-return-to-zero optical signals at 10 Gbps after propagation through up to 125 km optical fiber link. During the learning phase, a separation-loss function is optimized in order to maximally separate the transmitted levels of 0s from the 1s, which implies an optimization of the bit-error-rate. Testing of the NN device shows that the excess losses introduced by the NN device are compensated by the gain in transmitted signal equalization for a link longer than 100 km. The measured data are reproduced by a model which accounts for the optical link and the NN device. This allows simulating the network performances for higher data rates, where the device shows improvement with respect to the benchmark both in terms of performance as well as ease of use.
\end{abstract}

\section{Introduction}
Optical  fibers  are the backbone of Internet since they allow  data  transmission  at  large  bandwidths  and  long  distances. To increase the  capacity of the optical links large input optical power signals are needed to  compensate  for fiber  losses \cite{agrawalfiber}.  In these conditions  of  high-power  transmission,  both  linear  and  nonlinear  effects  alter  the  shape of the transmitted optical  pulses \cite{ip2008compensation}, which  implies  the  necessity  of  distortion  compensation  in the optical network. Nowadays,  signal recovery (equalization) is mostly accomplished  by  digital devices that introduce latency, delay and power consumption \cite{cartledge2017digital}. A clear example is observed in the trend to replace simple intensity modulation direct detection (IMDD) transceivers with more performing but costly and power-hungry coherent transceivers \cite{kikuchi2015fundamentals}, where digital signal processing (DSP) devices allow running algorithm to restore the data \cite{zhao2019advanced}. Different numerical approaches to correct for both linear and nonlinear optical fiber impairments exist with an emerging trend to use artificial intelligence-based algorithms \cite{mata2018artificial}.

To reduce the cost and power consumption of optical links, it is desirable to introduce equalization techniques also for simple IMDD systems. Even linear impairments, such as Chromatic Dispersion (CD), Polarization Mode Dispersion (PMD), Symbol Timing Offset and Optical filtering, severely distort the transmission \cite{agrawalfiber}. Among these impairments, one of the most severe is CD which causes a broadening of the optical pulse and the associated intersymbol interference \cite{elrefaie1988chromatic}. To compensate or correct for CD several types of equalization techniques have been introduced, among which dispersion compensated optical fiber and Bragg gratings are the most diffused ones \cite{agrawalfiber}. These are based on the use of dispersion-compensated units which recover the initial undispersed signal by counter-acting the CD effect. Another approach relays on the use of a dispersion compensating photonic-integrated programmable lattice filter formed by cascaded Mach-Zehnder interferometers \cite{brodnik2018extended}. An alternative to these approaches is the use of integrated photonic neural networks \cite{stark2020opportunities}. Their  advantages derive  from  operating  the  corrections  directly  in  the  optical  domain,  drastically  reducing  the  power  demand  and  the  latency, as well as in the flexibility of the equalization which can be learned directly on the deployed link and, therefore, can be easily adapted to optical link variations. Few hardware implementations of this concept exist  
\cite{katumba2019neuromorphic,ranzini2021experimental,ranzini2019tunable, argyris2018photonic, li2021micro,boshgazi2022virtual,li2019100}.\\

Here, we propose and validate the use of a small silicon photonics 4-channels delayed complex perceptron \cite{mancinelli2022photonic} to equalize a 10 Gbps IMDD 100 km long optical link. In the proposed photonic neural network (NN), the input signal is split into 4 channels where the combined actions of delay lines and tunable phase shifters  create the desired interference pattern at the output that counteracts the intersymbol interference. This  working principle has been applied  to  compensate for distortions induced by linear effects during  propagation  in a single-mode fiber. Equalization is performed on-chip and no external data processing is thus needed, except for the training phase. NN training  is based on a Particle  Swarm  Optimizer (PSO) \cite{schutte2004parallel}. Moreover, being the NN of the Feed Forward type \cite{mancinelli2022photonic}, the latency induced in signal processing is maximally reduced.

\section{Procedures}
\label{sec:procedures}
\begin{figure}[b]
    \centering
    \includegraphics{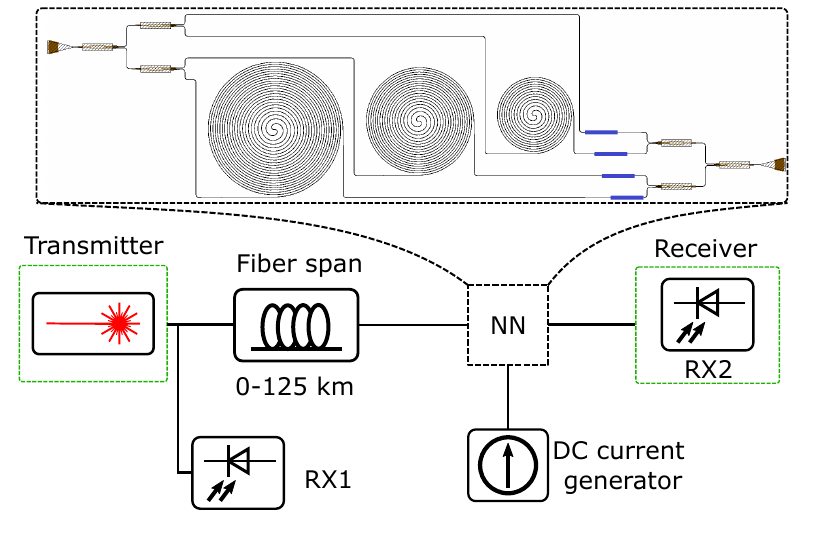}
    \caption{Experimental setup. The full link consists of a transmission stage, the optical fiber, the neural network (NN) device and the receiver stage. Two fast photodetectors (RX1 and RX2) allow for measuring the input and the transmitted signals. The inset shows the actual design of the NN device, where one can observe the cascaded 1x4 and 4x1 splitter and combiner, the three spirals, and the four phase shifters (small blue rectangles) connected to the external DC current controller. Details are given in Appendix \ref{sec:methods_exp_setup} or in \cite{mancinelli2022photonic}.}
    \label{fig:setup_simple}
\end{figure}

The small NN device, whose design is shown in Fig. \ref{fig:setup_simple}, is based on a delayed complex perceptron\cite{mancinelli2022photonic}. The input signal ($u(t)$, the input complex field) is split into four waveguides by a cascade of 1x2 multimode interferometers (MMIs). On each $k$-th waveguide but the first ($k=$1), a spiral forms a delay line which adds to the input signal a delay $\Delta_k=(k-1) \Delta_t$, where $\Delta_t=50 $ ps has been determined from the signal bitrate that, in this case, is 10 Gbps in NRZ (Non-Return-to-Zero) modulation. After the delay stage, the $k$-th waveguide hosts a delayed copy of the input $u(t)$, namely $u_k(t)=u(t-\Delta_t(k-1))$ with $k$=1, $\dots$, 4. Now, the signal undergoes a phase modulation performed by phase shifters realized with current-controlled heaters. In this way, the signals in each waveguide are weighted with $w_k=a_k\exp (i \phi_k)$ where $a_k$ stands for the spiral losses and $\phi_k$ for the added phase. After the weighting section, the four signals are recombined by means of a 1x4 combiner, realized using a cascade of 2x1 MMIs, which performs the operation $\sum_{k=1}^{4} u_k(t) w_k$. The output signal is then detected by a fast photodetector, which closes the processing by performing a nonlinear transformation, i.e. the detected signal intensity is $y(t)=\left|\sum_{k=1}^{4} u_k(t) w_k\right|^2$. 

\textcolor{black}{ 
The delayed complex perceptron acts as a 4-tap filter. The complexity of the layout in terms of the number of taps $N_T$ and delay unit $\Delta_t$ is determined in relation to the input bitrate $B$ and the target propagation distance $L$. This relation can be empirically described as
\begin{equation}
    N_T = \text{int} \left( \frac{1/B + \vert L \beta_2 \Delta \omega \vert }{\Delta_t} \right).
    \label{eq:ntaps}
\end{equation}
Here the numerator represents an estimate of the new pulse width, obtained as the sum of the initial bit time slot $1/B$ and the pulse broadening $\Delta T$ induced by CD on a gaussian pulse propagating in a fiber \cite{agrawalfiber}.  $\beta_2$ represents the Group Velocity Dispersion parameter, and $\Delta \omega$ the pulse bandwidth. Substituting in Eq. \ref{eq:ntaps} the parameters for the propagation of a 10 Gbps NRZ PRBS ($\Delta \omega \approx 2 \pi \times 10$ GHz) through a $L = 100$ km long standard SM G.652D fiber ($\beta_2 = -0.021$ ps$^2$/m), one obtains a pulse broadening of $\Delta T = 130$ ps, that for $\Delta_t = 50$ ps corresponds to $N_T \approx 4$. The choice of $\Delta_t$ is the result of a trade-off between a sufficient sampling of the information of a single-bit time slot at recombination (at least 2 samples per bit) and the aim of having a restricted number of channels to contain the excess losses.
}

The NN device is fabricated on a Silicon-on-Insulator (SOI) platform within a multi-project wafer run at IMEC-Belgium. The waveguides are 220 nm thick and have a width of 450 nm, allowing for single-mode operation on both polarizations at 1550 nm. The input and output gratings \textcolor{black}{ have a footprint of 50 $\mu$m $\times$ 30 $\mu$m and fix the polarization to the Transverse Electric (TE). The $1\times2$ and $2\times1$ MMIs used for splitting and recombination of the optical signal have a footprint of 20 $\mu$m $\times$ 100 $\mu$m. The gratings and the MMIs are implemented using the proprietary IMEC PDKs. Phase shifters are based on current-driven heaters realized as 60 $\mu$m-long and 0.6 $\mu$m-wide with a resistance of $60\;\Omega$ placed on top of an 800 nm thick Silica cladding}. $\Delta_k$ are realized with spirals of a length $k$-th multiple of 3.56 mm (corresponding to a delay of $\Delta_t=$ 50 ps). \textcolor{black}{The optical losses of the spirals due to surface roughness present on the waveguide and by the  bends in the curved optical paths have been measured to 6 dB/cm \cite{mancinelli2022photonic}. These result in an attenuation of 2.1 dB, 4.3 dB, and 6.4 dB for $k = 2,\dots,4$, respectively.} The NN device's insertion losses have been estimated to be 8.2 dB at 1550 nm. The chip is placed on a Proportional-Integral-Derivative (PID) controlled Peltier cell that keeps its temperature at 21$^{\circ}$C.\\
The experimental setup is represented in Fig. \ref{fig:setup_simple}. In the transmission stage, a tunable laser source (TLS) operating at 1550 nm is modulated as a NRZ 10 Gbps Pseudo-Random binary sequence (PRBS) of order 10 and period $2^{10}$ bits. A 50:50 Fiber Optic Splitter sends half of the signal to a fast photodiode (RX1). The other half is coupled to an optical fiber span, where distortions induced by CD are accumulated. The length of the span goes from a minimum of 0 km to a maximum of 125 km, with a granularity of 25 km. The distorted signal enters the NN device for optical processing. DC current controllers set the currents in the heaters. The output signal from the NN device is coupled to a fast photodiode (RX2) at the receiver stage. Both RX1 and RX2, which monitor $y_{in}(t)$ and $y_{out}(t)$, respectively, are connected to a 40 GSa/s oscilloscope with a 16 GHz bandwidth.  The Signal-To-Noise ratio (SNR) at the receiver RX2 can be varied by using a Variable Optical Attenuator (VOA2) inserted after the NN device (see Appendix \ref{sec:methods_exp_setup}). 

For each measure, the DC controller sends pre-set currents to the NN device and a triggering signal to the oscilloscope. The acquisition is delayed by 1 ms from the arrival of the triggering signal to let the optical signal stabilize at the output of the NN device, according to the thermal relaxation time of the heaters (few tens of $\mu$s). The observation window of the oscilloscope is 1 $\mu$s wide, allowing the observation of at least 9 periods of PRBS at each acquisition. Four samples per bit are available because of the 40 GSa/s sampling rate. In what follows, the samples in each bit are labeled from 1 to 4, the 4$^{\text{th}}$ being the most recent. Acquired sequences are then under-sampled, obtaining a sub-sequence constituted by $n$-th sample in each bit of the full trace and being the chosen sample the most representative of the actual bit value (typically the closest to the center of the bit). An operation performed over the under-sampled sequence at the $n$-th sample in each bit will be mentioned as performed over the $n$-th sample.  \\

Input (from RX1) and output (from RX2) signals are aligned by exploiting their cross-correlation, obtaining $y_{in}$ and $y_{out}$. The under-sampling at the $n$-th sample of the two sequences provides $\overline{y}_{in}$ and $\overline{y}_{out}$, respectively. The output signal $\overline{y}_{out}$ is compared with $\overline{y}_{in}$ to label the 1 level ($\overline{y}_{out,H}$) or 0 level ($\overline{y}_{out,L}$).

NN training procedure is performed off-chip using fully automatized software. An analog loss function $\mathcal{L}$ is created to obtain the largest possible separation between the distributions of signal levels expected as 1s or 0s in the output signal. This quickly minimizes the associated bit error rate (BER) since it is directly linked to the overlap between these two distributions. Indeed, in presence of random gaussian noise characterized by standard deviations $\sigma_0$ and $\sigma_1$ affecting 1s and 0s in the bit sequence, the BER can be computed as \cite{agrawalfiber}
\begin{equation}
    \text{BER} = \frac{1}{4} \left[ \text{erfc}\left( \frac{I_1 - I_D}{\sigma_1 \sqrt{2}} \right) + \text{erfc} \left( \frac{I_D - I_0}{\sigma_0 \sqrt{2}} \right) \right] .
    \label{eq:ber_erfc}
\end{equation}
Here $I_{(0,1)} = \langle \overline{y}_{out,(L,H)} \rangle$ are the average levels for 1s and 0s, $\sigma_{(0,1)}$ are their standard deviations, $I_D$ is the decision threshold, and erfc is the complementary error function. A BER reduction can thus be obtained by maximizing $I_1 - I_0$. 
$\mathcal{L}$ measures the spacing between the tails of the distributions related to $\overline{y}_{out,H}$ and $\overline{y}_{out,L}$. The training's goal is the maximization of this spacing, therefore we call it the separation loss function. Having $\overline{y}_{out,(L,H)}^{i}$ the measured signal values in a sequence, the separation loss function is expressed as 
\begin{equation}
    \mathcal{L} = E[0] - E[1] = \frac{1}{N_L} \left[
     \sum_{i = 1}^{N_{L}} \overline{y}_{out,L}^{i}
    \right]
    -
    \frac{1}{N_H}
    \left[ \sum_{i = 1}^{N_{H}} \overline{y}_{out,H}^{i}
    \right],
\end{equation}

$E[0]$ and $E[1]$ are estimates of the tail position in the two distributions. In $E[0]$, $i$ runs over the samples such that $\overline{y}_{out,L}^{i} > I_0 + 1.28 \sigma_0$, namely $\overline{y}_{out,L}^{i}$ is part of the group of the rightmost $N_L$ points corresponding to the 10\% of the population of the $\overline{y}_{out,L}$ distribution. Similarly, $\overline{y}_{out,H}^{i} < I_1 - 1.28 \sigma_1$ is part of the group of the leftmost $N_H$ points corresponding to the 10\% of the population of $\overline{y}_{out,H}$ distribution. The PSO is adopted for training \cite{schutte2004parallel}, which is performed in a condition of no attenuation in front of RX2, i.e. an average optical power at RX2 of about 0 dBm and a SNR of 11.2 dB. 

\textcolor{black}{In light of the differentiability of $\mathcal{L}$ with respect to the currents controlling the induced phase shifts in the device, other choices for the training algorithm are possible, including a Back-Propagation (BP) technique. During the experimental phase, we performed some tests using an adapted version of the Adam algorithm \cite{kingma2014adam}, which is a gradient-based alternative in which the descent proceeds with a memory of the previous iterations. Such weighted adaptation of the gradient is well known to make more robust the trajectory towards the local minimum in the presence of noise and is often preferred versus the standard BP algorithm. The algorithm proved to be more time-efficient but possibly limited by premature endings of the research at a local minimum. Therefore, here we chose to rely on the PSO, which guaranteed the robustness and the repeatability of the final outcomes.}

After the training phase, the testing phase is performed via a scan over the power at the receiver (PRX) made by varying the attenuation of VOA2 in front of RX2, which corresponds to a scan over the SNR at the RX2. For each PRX value, 50 acquisitions for a total of $5 \times 10^5$ bits with the trained currents set are performed, evaluating the BER for each measure. The BER is defined here as the cumulative error between the digitized input and output signals. The digitized signals are obtained by applying a threshold to $\overline{y}_{in}$ and $\overline{y}_{out}$. At each evaluation, the optimal sample for the generation of $\overline{y}_{out}$ and the optimal threshold which minimizes the BER are selected. The threshold is chosen among 10 possible equally spaced levels between the minimum and maximum of the signal. Training and testing procedures are performed for multiple lengths of the fiber span and then compared with the corresponding reference curves obtained without the NN device. \\

The full optical link (from the transmission to the receiver stages) is simulated to model the effect of the NN device (see Appendix \ref{sec:comput_model} for details). Also in the simulation, the NN's training is performed by optimizing the separation loss function with the PSO. Noise is added as described in the Appendix \ref{sec:noiseModel}. The sampling of the oscilloscope is modeled as well. BER is computed as in the experimental case.

\section{Results}

\begin{figure}[t!]
    \centering
    \includegraphics[width=\textwidth]{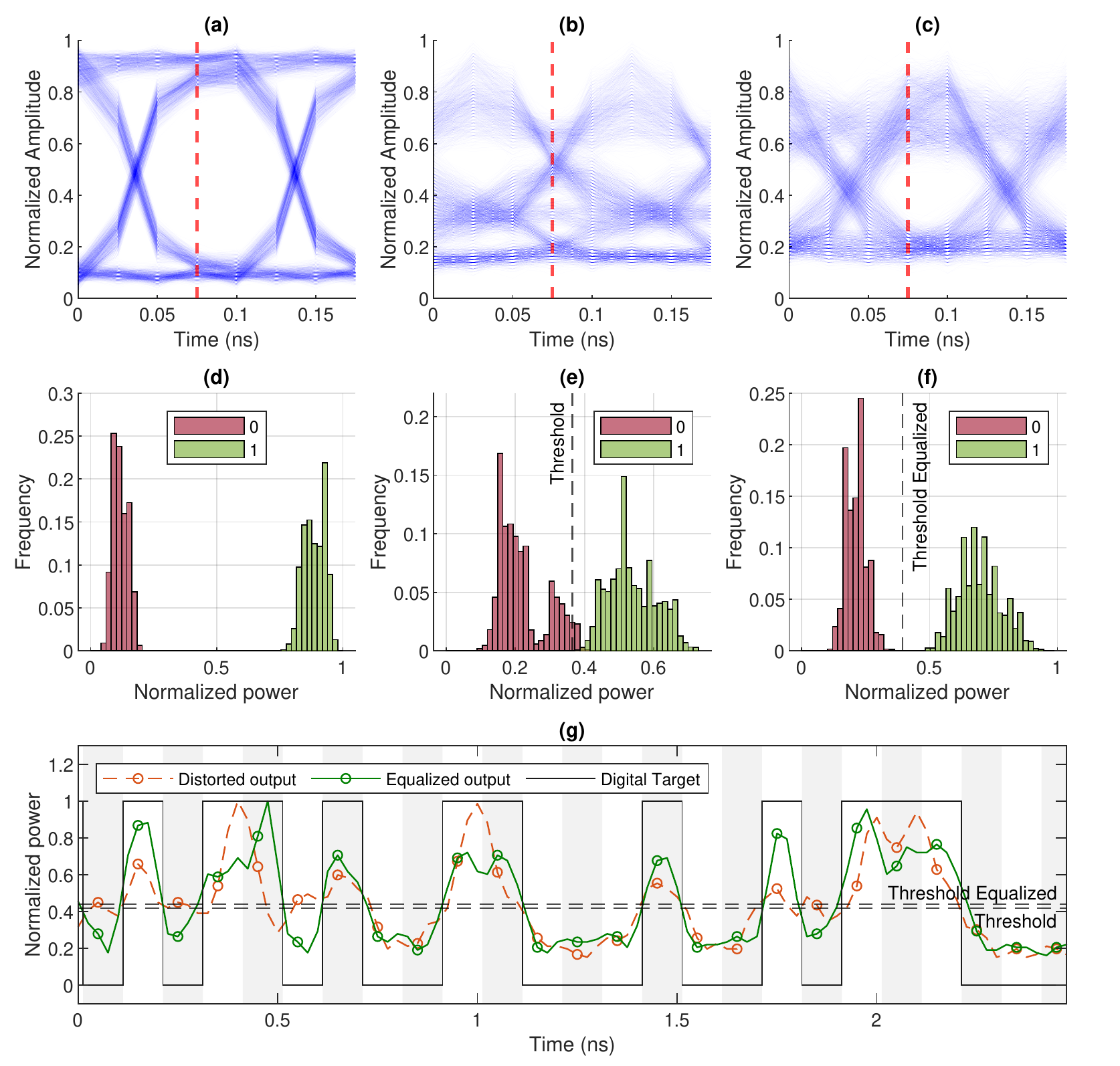}
    \caption{Testing of the neural network device on a 125 km fiber link. \textcolor{black}{(a-c) Eye diagrams of (a) the input signal, (b) the transmitted signal without the NN device in the link (c) and the transmitted signal with the trained NN device in the link. The dashed red lines refer to the sampling time that is used to generate the histograms.} (d-f) Histograms of the power levels associated with the 2$^{\text{nd}}$ sample in the input bit sequence for (d) the input signal, (e) the transmitted signal without the NN device in the link and (f) the transmitted signal with the trained NN device in the link. The red and green columns refer to the input 0s and 1s, respectively. The dashed vertical lines set the decision threshold. The sub-optimal choice of the threshold in (e) derives from a discretization of the possible threshold values. (g) Actual temporal sequences for the transmitted signal as recorded by the RX2 (line with the NN device in the link, dashed without). The light black line refers to the digital input. Circles highlight the 2$^{\text{nd}}$ sample in each bit that is used in the BER calculations. Horizontal dashed lines show the different thresholds for the equalized and non-equalized data used to minimize the BER. Threshold values are rescaled according to the normalization used to plot the curves.}
    \label{fig:histograms}
\end{figure}

The equalization effect of the NN device for a span of 125 km is summarized in Fig. \ref{fig:histograms}. \textcolor{black}{The eye diagrams in panels (a-c) show the three aperture conditions reached after the modulation at the transmitter (a), after the fiber propagation (b) and after the equalization performed by the NN device (c), respectively. CD generates a closure of the eye diagram, as a consequence of the intersymbol interference. Particularly evident in Fig. \ref{fig:histograms}(b) is a high-density region between normalized amplitude values of 0.3 and 0.4 \textcolor{black}{crossed by the red-dashed line}, which represents a raise of the zero-level induced by the interference of a low bit with neighboring bits in the high state. The action of the NN device partially restores the aperture (panel (c)), eliminating the intermediate level seen in panel (b).} The same scenario is presented in panels (d-f), where the histograms report the distributions of the optical power levels expected as 0s or 1s associated with the $2^{\text{nd}}$ sample in the bit, in the input (d), non-corrected (e) and corrected (f) output signals, respectively. An example of their time evolution is reported in Fig. \ref{fig:histograms}(g) with normalized amplitudes. Data are collected with an SNR = 11.2 dB at RX2. In this regime, the evaluation of the BER is not limited by the SNR, but by the fiber dispersion that generates intersymbol interference. The distorted output in Fig. \ref{fig:histograms}(g) shows clearly the presence of pulse broadening and the consequent generation of an intersymbol. Bits expected as 0s preceded or followed by a 1 are raised close to the 1s, increasing thus the probability for errors. As a consequence, the distributions of power levels for 0s and 1s widen and the gap between the distributions reduces, as shown in Fig. \ref{fig:histograms}(e). This leads to an increased BER. As clear from Fig. \ref{fig:histograms}(f), the corrective action of the NN device partially restores the two distorted distributions of Fig. \ref{fig:histograms}(e), thus decreasing the BER.\\

\begin{figure}[t]
    \centering
    \includegraphics[width=\textwidth]{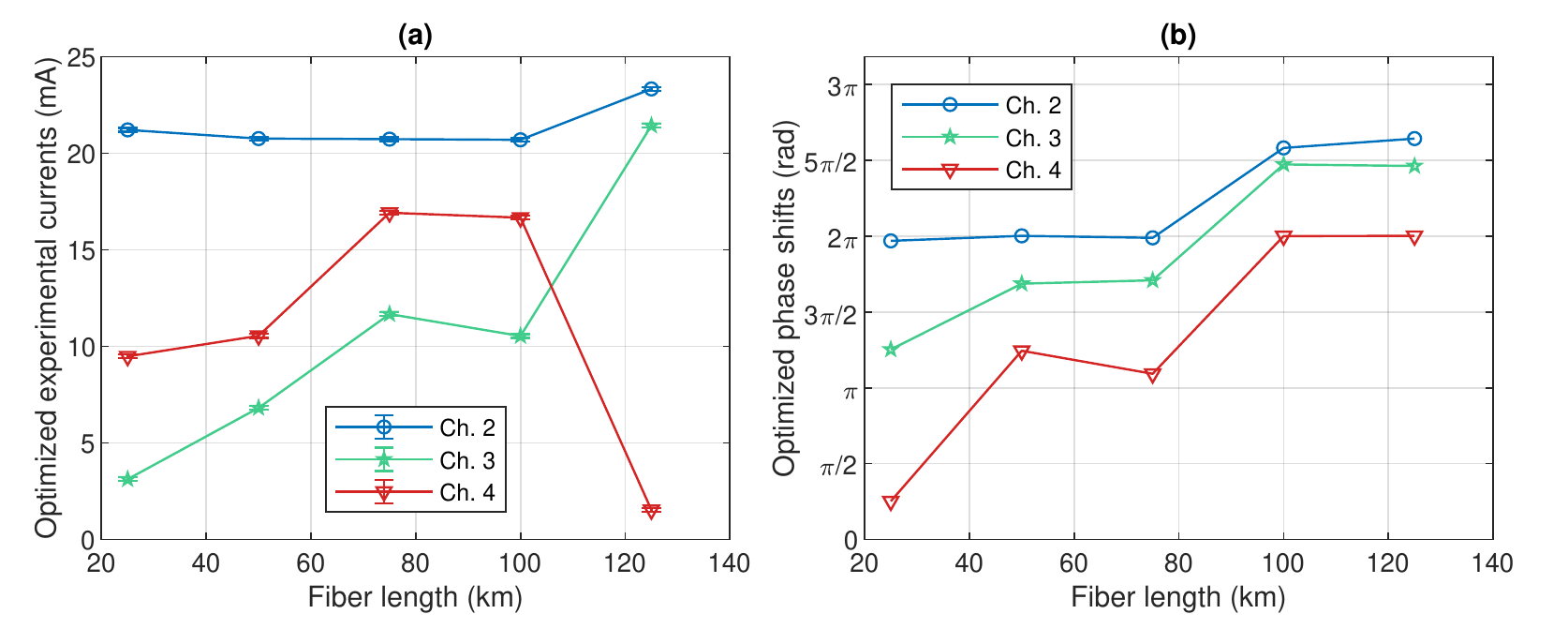}
    \caption{\textcolor{black}{Training outcomes for signal equalization as a function of the fiber link length. Optimized experimental currents (a) and simulated optimized relative phase shifts (b) in channels 2 (blue dots), 3 (green stars), and 4 (red triangles) after the training. Error bars in (a) (barely visible) derive from the instrument output precision, not from statistics. The phase shifts in (b) are measured in each channel with respect to the 1$^{\text{st}}$ channel (no spiral). PSO is chosen as the training algorithm in both cases.}}
    \label{fig:output_phases}
\end{figure}

\textcolor{black}{The training returns a set of 3 optimal currents, associated with the channels in the NN device. One can then model the relative recombination phase shift $\phi_k$  (with $k=2,\dots,4$) used for the weight $w_k$ in the $k$-th channel  with respect to the first channel (chosen as reference) as
\begin{equation}
    \phi_k = \phi_k^o + i_k^2 \gamma_k ,
\end{equation}
where $\phi_k^0$ is the relative phase measured at zero currents, $i_k$ is the optimal current in the $k$-th channel and $\gamma_k$ is the conversion factor between the dissipated thermal power in a resistor and the induced phase shift in the underneath waveguide. Measurements conducted on test resistor structures yield $\gamma_k \approx 0.01$ rad/mA$^2$ \cite{biasi2022thermal}, while finding $\phi_k^o$ is cumbersome due to the uncertainties in the optical path lengths and widths caused by the finite fabrication resolution. Therefore, for the sake of clarity, we show in Fig. \ref{fig:output_phases}(a) the currents used for the trained NN at different fiber lengths, and in Fig. \ref{fig:output_phases}(b) the corresponding phase shifts obtained from the simulation. In panel (b) it appears that longer optical links require an increase of the phase shift to about 2$\pi$ in each channel, meaning that the delayed copies constructively contribute to the output \cite{mancinelli2022photonic}. Thus, the NN device weights more the contributions by the longer delay lines since a larger pulse broadening is to be compensated. \\}

\begin{figure}
    \centering
    \includegraphics[width=\textwidth]{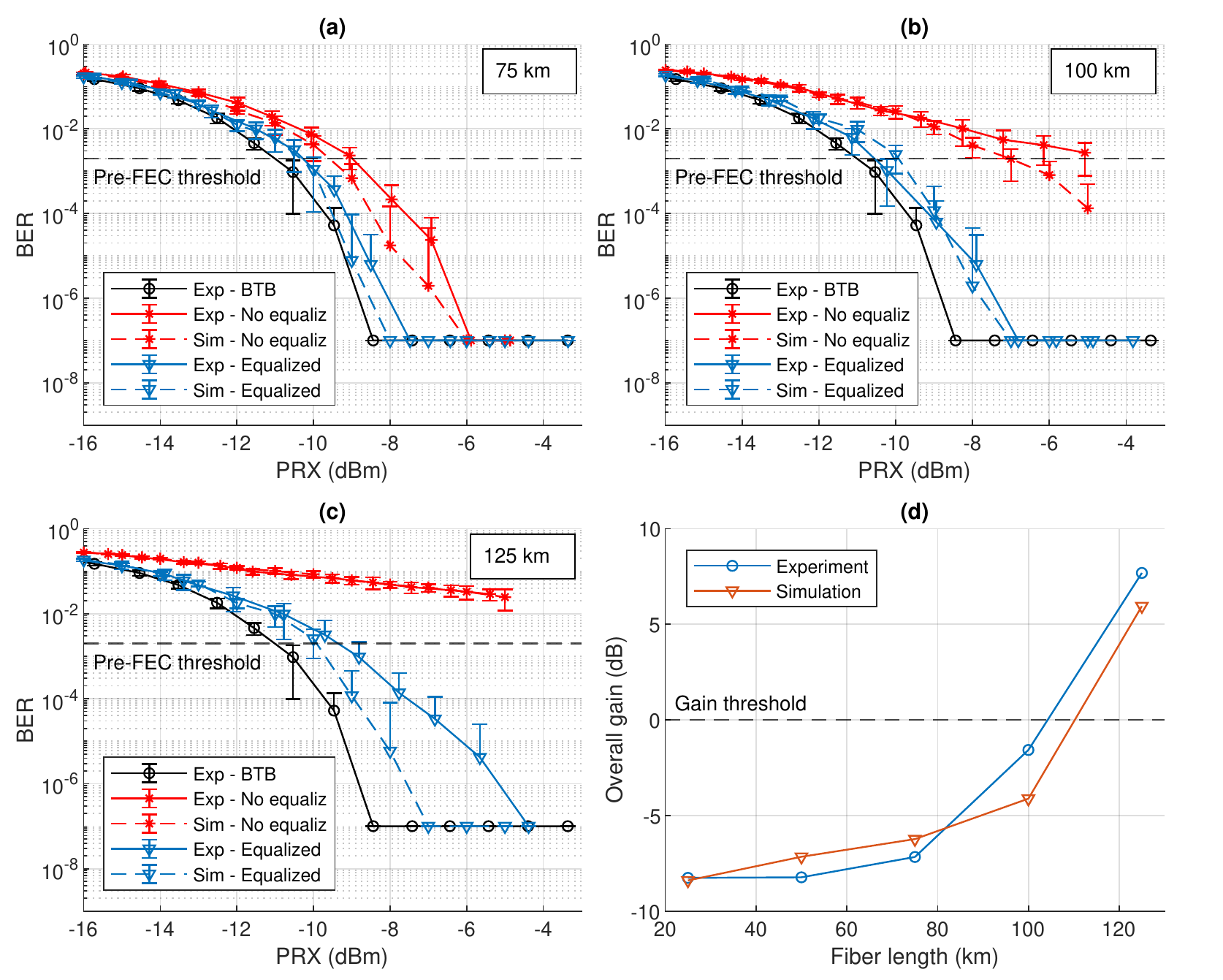}
    \caption{(a-c) Experimental (full lines) and simulated (dashed lines) BER versus PRX curves: black discs refer to the back-to-back (BTB) configuration (the transmission stage is directly interfaced to the receiver), the red stars to the transmission by a fiber link, the blue triangles to the transmission by a fiber link with the NN device. Dashed horizontal black lines refer to the pre-FEC threshold value. Error bars are calculated as the standard deviation over multiple acquisitions (see Appendix \ref{sec:methods_exp_setup}). Error-free points are replaced by $1\times10^{-7}$ due to the finite dimension of the data set. The used fiber link is 75 km long (a), 100 km long (b) or 125 km long (c). (d) Overall gain provided by the NN device as a function of the fiber link length (blue discs experiment, orange triangles simulation). Gain is given as the improvement of the PRX at $\text{BER}=2\times 10^{-3}$ when the NN device is used with respect to results without the NN device. The dashed line marks the threshold above which the gain guaranteed by the equalization is greater than the NN device excess loss of about 8.5 dB.}
    \label{fig:bervsprx_fom}
\end{figure}

Figures \ref{fig:bervsprx_fom} (a,b,c) report the simulated and experimental BER versus PRX profiles obtained for different fiber spans. The Back-to-Back (BTB) configuration (namely with no NN device and no fiber, black curve) measures the TX/RX performance. For low PRX values (namely low SNR), BER is dominated by noise, which is present in the output signal at the receiver regardless of the length of the fiber. Thus, all the profiles overlap in this region. On the contrary, for higher PRX, the SNR increases too and the most dominant contribution in the BER is provided by distortions in the signal induced by cumulated chromatic dispersion. These distortions become more important for longer fibers, causing a worsening of the BER even at high PRX. In fact, the dispersion length for this system is $L_D = T_0^2 / \vert \beta_2 \vert = 2 \pi c_0 T_0^2 / ( \lambda^2 \vert D  \vert) = 450$ km, using $D = 17.2$ ps/nm/km and $T_0=100$ ps. The effects of the corrections operated by the NN device are evident for long fiber lengths ($\geq$ 100 km)  when the amount of distortion to be compensated is significant. The NN device almost recovers the BER versus PRX curves to the reference optimal case (BTB).\\

The gain brought in by the action of the NN device can be described starting from the PRX values corresponding to the same BER in the experimental curves. The reference BER value is considered to be $2\times 10^{-3}$, being this a typical BER-threshold value for Forward Error Correction (pre-FEC threshold) \cite{Mizuochi}. The corresponding PRX values are interpolated for each BER versus PRX profile obtained both with and without the NN device, producing respectively $PRX(w)$ and $PRX(w/)$. Figure \ref{fig:bervsprx_fom}(d) reports the corresponding experimental and simulated overall gain obtained as $PRX(w) - PRX(w/)$ subtracted with the Excess Loss (EL) introduced by the trained NN device. Note that the EL depends on the actual weights configuration since the output signal results from the interference of the weighted and delayed copies of the input \cite{mancinelli2022photonic}. We use the best-case scenario and we neglect in the EL calculations the 8.2 dB contribution of the grating losses. Note also that the values of $PRX(w/)$ for 100 km and 125 km fiber spans are extrapolated from the corresponding BER versus PRX since no data at the pre-FEC threshold are available. The horizontal line at the null value highlights the point where the gain generated by the NN device compensates for its excess loss. This happens for fiber lengths above 100 km for the used bit rate. The NN device has to be considered as underperforming, being the 6 dB/cm spiral propagation losses unusually higher than the expected nominal value of 2 dB/cm for IMEC processing \cite{absil2015imec}. Improvements in the fabrication could increase further the performance of this already working NN device. \\

\begin{figure}
	\centering
	\includegraphics[width=\textwidth]{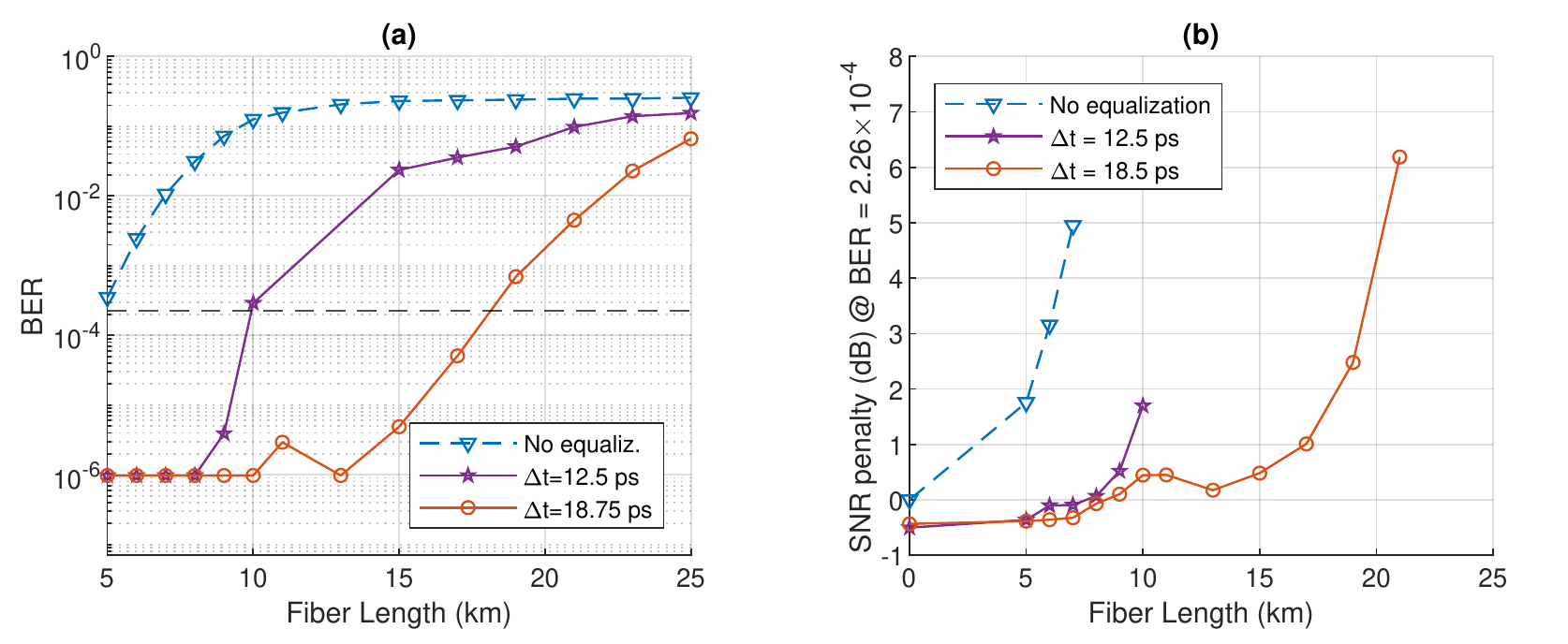}
	\caption{(a) BER versus link length with SNR$=$12 dB at the receiver for the link without the NN device (dashed blue line and triangles), with the NN device and a delay granularity of 12.5 ps (full purple line and stars), and with the NN device and a delay granularity of 18.75 ps (full orange line and circles). (b) SNR penalty at a BER of $2.26\times 10^{-4}$ as a function of the link length without the NN device (dashed blue line and triangles), with the NN device and a delay granularity of 12.5 ps (full purple line and stars), and with the NN device and a delay granularity of 18.75 ps (full orange line and circles). Penalty is calculated from the back-to-back performance. Curves are interrupted at the last fiber length value for which it was possible to interpolate the chosen BER threshold in the corresponding BER versus PRX profile. }
	\label{fig:performance_comparison}
\end{figure}

\section{Conclusions}
The model of the NN device allows accessing working conditions that are not explorable with the present integrated version of the NN device. Indeed, its versatility is limited by the fixed delay lines which are set for a 10 Gbps data rate. On the contrary, the simulations allow adopting a higher modulation frequency by adapting the delay lines to different bit rates. We can thus compare the performance of our NN device with the results obtained in \cite{katumba2019neuromorphic} and \cite{ranzini2019tunable} which can be considered as a benchmark for the current state of the art for short reach (up to 25 km) access link applications. The first approach\cite{katumba2019neuromorphic} is based on the reservoir computing paradigm where a photonic integrated circuit composed of delay lines and beam splitters arranged in a swirl topology forms the reservoir. The second approach\cite{ranzini2019tunable} is based on the spectral decomposition technique where the spectral content of the optical carrier is divided into slices and analyzed following an all-optical/hybrid approach.

The comparison starts by tuning the parameters of our simulation in order to reproduce the BER versus fiber length  profile of \cite{katumba2019neuromorphic} at 40 Gbps. In particular, the SNR has been fixed to 12 dB and each BER value is obtained as an average over $1.024 \times 10^6$ transmitted bits. The parameters are kept unaltered for the other runs too, including the training and subsequent testing phase for the NN device. The NN has been modeled with delay lines introducing a shift of half (12.5 ps) or three-quarters (18.75 ps) of a bit. To compare our NN device with the ones of \cite{katumba2019neuromorphic} and \cite{ranzini2019tunable} we used the same representative performances as in these works. First, the BER as a function of the link length for the NN device at 40 Gbps is reported in Fig. \ref{fig:performance_comparison} (a). A clear BER improvement is observed where the equalization provided by the trained NN device ensures an extension of the link reach up to almost 20 km when a delay of 18.75 ps is used. A comparison with the results in \cite{katumba2019neuromorphic}  shows that the present NN device provides better BER performances up to 20 km fiber length. Despite its simplicity, the NN device outperforms the swirl-based reservoir without the need for electrical data post-processing \cite{katumba2019neuromorphic}.

In \cite{ranzini2019tunable}, the SNR penalty is used as a figure of merit. This is defined as the increase in the SNR needed to achieve the same BER as that of the BTB configuration and calculated at the pre-FEC threshold of $2.26\times 10^{-4}$. The SNR penalty of our trained NN device for an NRZ 40 Gbps data rate as a function of the fiber link length is shown in Fig. \ref{fig:performance_comparison}(b). When the NN device is used with a delay line granularity of 18.75 ps, the SNR penalty stays below 1 dB up to a link length of 18 km. Compared to the performances of the devices discussed in \cite{ranzini2019tunable}, the present NN device is doing better than the 1-stage and 2-stage fully optical devices but worse than the 4-stage fully optical devices, which has however a significantly larger complexity (it requires 30  Mach-Zehnder interferometers) than our NN design.

\textcolor{black}{
The performances of our device validate its use for signal equalization, suggesting further studies for the optimization of the layout for in-line applications. We foresee next-generation devices equipped with an augmented number of channels and amplitude modulators in each tap to allow for much larger adaptability to the different transmission scenarios (bitrate, modulation format…). A transceiver-packaged version of these optimized devices would provide significant advantages even at high modulation frequencies (up to 100 Gbps) at metro propagation distances (up to 100 km). These in-line transceivers relieve the computational efforts of complex DSPs both in coherent and IMDD systems, in addition to a latency reduction. Most important for short-reach applications is a significant reduction in power consumption, which for the present NN accounts for 70 mW, to be compared with the typical $>$ 1 W for DSP (Table 19.1 in \cite{Kilper2020energy}). Thus, simplified DSPs (e.g. less power-hungry) will be required to achieve the same BER over longer distances without reducing the carrier frequency. 
}

In summary, we demonstrated a simple concept of a feed-forward neural network device that is able to correct linear signal distortion both on a metro network (10 Gbps, 100 km) and on a high-speed short-reach access link (40 Gbps, 20 km). For different applications which have different data rates, proper tuning of the nodes' delays is needed.

\section*{Appendix}

\appendix

\section{Experimental setup}
\label{sec:methods_exp_setup}

\begin{figure}[t]
    \centering
    \includegraphics[width=\textwidth]{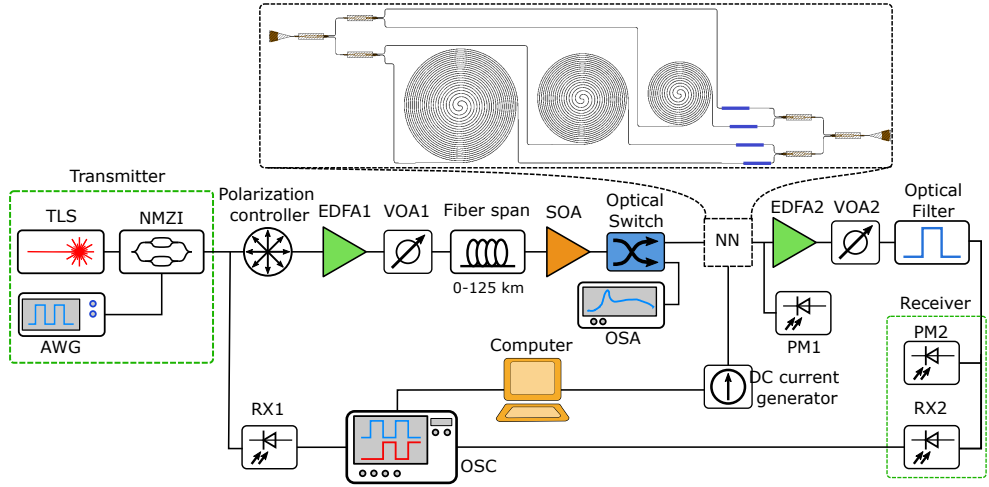}
    \caption{Experimental setup. The different symbols are self-explanatory and are discussed in the text. The inset shows the design of the NN device.}
    \label{fig:setup}
\end{figure}

Figure \ref{fig:setup} presents the experimental setup. The tunable laser source (TLS) is constituted by an InGaAs-based semiconductor laser that can be thermally tuned around 1550 nm. The source is modulated by a Nested Mach-Zehnder interferometer (NMZI) \textcolor{black}{with 30 GHz of electro-optical bandwidth} and driven by an Arbitrary Waveform Generator (AWG) with 30 GHz of electrical bandwidth and a sampling rate of up 64 GSa/s. After the modulation stage, a Fiber Optic Coupler with 50\% coupling ratio addresses half of the optical signal to a fast photodiode (RX1\textcolor{black}{, 20 GHz bandwidth}) which detects the input signal. A polarization controller allows tuning the local compression and torque applied to the fiber itself, inducing a polarization change.
    
An Erbium-Doped Fiber Amplifier (EDFA1) amplifies the optical signal to 20 dBm and then a Variable Optical Attenuator (VOA1) controls the effective power level launched into the fiber link made of an SM G.652D fiber with a nominal loss coefficient of 0.2 dB/km. The fiber link length is varied during the experiments from a minimum of 0 km to a maximum of 125 km with a granularity of 25 km. A Semiconductor Optical Amplifier (SOA) with a small signal gain of 13.4 dB is inserted at the end of the fiber link to partially recover the fiber link attenuation. The amplified optical signal is then sent to a switch that allows addressing the optical signal respectively to an Optical Spectrum Analyzer (OSA) or to the input grating of the NN device. Here the optical signal is processed by the NN and, via the output grating coupler, is coupled to the output fiber. Currents sent to the NN device are provided by a terminal-controlled DC current generator. 
    
The output fiber is connected to a Fiber Optic Coupler with a $99.9:0.1$ coupling ratio to address 0.1\% of the optical signal to a Power Monitor (PM1). The other 99.9\% is sent to a second EDFA (EDFA2) (small-signal gain of 30 dB) followed by a second VOA (VOA2). The combined action of these last two elements regulates the amount of optical power detected to a level below the damage threshold of the fast photodiode RX2. Then, a tunable optical filter with 30 GHz bandwidth and 5 dB of insertion loss cleans up the signal from the out-of-band amplified spontaneous emission noise added by the amplification stages. Another Fiber Optic coupler with $99.9:0.1$ coupling splits the signal towards a second Power Monitor (PM2) and to another fast photodiode (RX2\textcolor{black}{, 20 GHz bandwidth}) which measures the output signal. Both RX1 and RX2 fast photodiodes are connected to a 40 GSa/s oscilloscope (OSC) with a 16 GHz bandwidth. \textcolor{black}{The bandwidth limit of the transmission line is thus fixed by the oscilloscope, having the narrowest bandwidth in the line.} \\

For each measure, the DC current generator sends pre-set currents to the NN device and a triggering signal to the oscilloscope. The acquisition is delayed by 1 ms from the arrival of the triggering signal to stabilize the NN device response, according to the thermal relaxation time of the heaters. The observation window of the oscilloscope is 1 $\mu$s wide, allowing the observation of at least 9 periods of PRBS at each acquisition. Four samples per bit are available because of the 40 GSa/s sampling rate. Each point in the BER versus PRX profiles (as those in Fig. \ref{fig:bervsprx_fom}) is obtained as an average over $N=100$ measurements, allowing getting a minimum non-null measurable value of 1/($N \times$ Number of bits in the sequence). Error bars are obtained as the standard deviation of the measured BER values for that point.

\section{Modeling of the experiment}
\label{sec:comput_model}

The main elements of the optical link have been modeled, simulating the modulation apparatus, the structure of the NN device, the propagation of the optical PRBS sequence in the fiber link and the noise contributions at the receiver. In the following, the reported numerical values for the parameters refer to the simulation performed at 10 Gbps (40 Gbps).\\

The model includes a Mach-Zehnder modulator driven by an electrical signal to imprint on the optical carrier a PRBS sequence of order 10 and period $2^{10}$ bits with an analog bandwidth of 30 GHz, an extinction ratio of 13.9 dB and a null chirp. \textcolor{black}{The sampling frequency of the electrical signal is fixed to 320 GSa/s, in order to preserve the information over a sufficiently large frequency range. The same sampling frequency is thus maintained also in the resulting modulated optical signal propagating across the simulated setup until the final detection process and simulated oscilloscope, which reduces the sampling frequency to 40 (160) GSa/s.}

The evolution of the signal is simulated by solving the linear Schr\"odinger equation in the Fourier domain. This can be derived following the approach proposed in \cite{agrawalfiber}, which reduces to 
\begin{equation}
    \Tilde{A}(z, \omega) = \exp\left[ i z \frac{\beta_2}{2} \omega^2 - \frac{\alpha}{2}z  \right]  \Tilde{A}(0, \omega)
    \label{eq:propEqFFT}
\end{equation}
where $\Tilde{A}(z, \omega)$ is the Fourier Transform of the temporal optical field envelope, $z$ is the propagation distance, $\beta_2$ is the group-velocity dispersion (GVD) parameter and $\alpha$ stands for the fiber losses. The result of this operation is the propagated temporal optical field envelope $A(z,t)$, obtained by applying the inverse Fourier transform to $\Tilde{A}(z,\omega)$. 

This complex optical field signal is then provided as input to a model of the NN device. The model simulates the action of 4 delay lines, each of them associated with a fixed attenuation value measured for the actual spiral length in the NN device. A tunable phase shift is applied to each channel before the output combiner, to simulate the action of the heater-actuated phase shifters. Note that Eq.\,\ref{eq:propEqFFT} was not used to simulate the signal propagation inside the NN device, since the dispersion length $L_D \approx 2$ km \cite{agrawalfiber} associated with the spirals is much longer than the length of the spirals themselves ( $\sim$ 1 cm). Effects deriving from chromatic dispersion can thus be neglected inside the NN device. \textcolor{black}{Therefore, the relative delay between the 4 channels has been emulated by inserting a shift of the proper amount of samples between the 4 sequences.}

After the combiner, the complex optical field signal is converted into the detected optical power (output signal) through the modulus square operation and, then, treated to account for the noise measured experimentally at the receiver (noise modeling is discussed in the next section). A band-pass filter with a bandwidth of 16 GHz (28 GHz) obtained with a $5^{\text{th}}$-grade Bessel polynomial is then applied to the detected output signal, simulating the electronic bandwidth of the oscilloscope. \textcolor{black}{An 8-bit vertical sampling with 100 mV full-scale is then applied to the output signal, together with a 40 GSa/s (160 GSa/s) horizontal sampling}. For each simulated acquisition, the position of the first sample in the first bit is randomly chosen in the first quarter of the duration of the bit itself. Indeed, in the experimental setup, the triggering signal for the oscilloscope comes from the DC generator which controls the phase shifters too. The oscilloscope is then asynchronous with the AWG and, therefore, the position of the first sample in the sequence is different in each acquisition. Depending on where the first sample falls in each bit, the contrast level in the acquired curve changes, possibly leading to a different BER result. 

The sampled output signal is then compared with the input signal according to the same modalities described in Section \ref{sec:procedures} for the real experiment. The only difference regards the training phase, during which the loss function is always evaluated at the $3^{\text{rd}}$ sample, being this close to the center of the bit and distant from the transients. Different fiber length scenarios are simulated using the PSO training algorithm. \\

After each run, the BER versus PRX curves are calculated. Each BER value appearing in the profiles is obtained as an average over $N=1000$ measurements, corresponding to a minimum non-null measurable value of 1/($N \times$ Number of bits in the sequence).

\section{Noise modeling}
\label{sec:noiseModel}

In the experimental setup, the optical amplifiers (EDFAs and SOA) act as noise sources, but the presence of the 30 GHz band-pass optical filter reduces their impact in deteriorating the SNR at the receiver. In the studied configuration, their contribution is negligible with respect to that introduced by the fast photodiode (RX2, receiver). The fluctuations in its response to the input optical power can be modeled as follows \cite{agrawalfiber}
\begin{equation*}
    \sigma^2 = \langle (\Delta I)^2 \rangle = \sigma_s^2 + \sigma_T^2 = 2q(I_p + I_d) \Delta f + (4 k_B T/ R_L) F_n \Delta f .
\end{equation*}
The first term accounts for the contribution coming from shot noise, being $q$ the electron charge, $I_p$ the average current, $I_d$ the dark current and $\Delta f$ representing the effective noise bandwidth of the detector. The second term describes fluctuations induced by thermal noise, being $k_B$ the Boltzmann constant, $T$ the temperature, $R_L$ the load resistor of the detector and $F_n$ the noise figure of its internal amplifier. For the current experimental setup, the previous equation becomes $\sigma^2 = \langle (\Delta V)^2 \rangle = m V_{meas} + q$ where $V_{meas}$ is the measured voltage at the oscilloscope, $m$ accounts for the proportional term due to shot noise and $q$ includes the noise contributions deriving from thermal noise and shot noise associated with the dark current. A characterization of the setup provided us with $m=0.0189$ mV and $q=0.2263$ mV$^2$.

\begin{backmatter}
\bmsection{Funding}
European Research Council (ERC) under the European Union’s Horizon 2020 research and innovation programme (grant agreement No 788793, BACKUP and No 963463, ALPI).

\bmsection{Acknowledgments}
We acknowledge a fruitful discussion with Stefano Biasi. This project has received funding from the European Research Council (ERC) under the European Union’s Horizon 2020 research and innovation programme (grant agreement No 788793, BACKUP and No 963463, ALPI). 

\bmsection{Disclosures}
M.M., P.B. and L.P. have filed a patent on the technology here described.

\bmsection{Data Availability Statement}
The data that support the findings of this study are available from the corresponding author upon reasonable request.

\end{backmatter}


\bibliography{sample}

\end{document}